\documentclass [twocolumn,aps,prb]{revtex4}
\usepackage{epsfig,graphicx,color,amssymb,amsmath,setspace}
\begin{document}
\title{Meir-Wingreen formula for heat transport in a spin-boson nanojunction model}
\author{Kirill A. Velizhanin\footnote{electronic mail: kirill@lanl.gov}}
\affiliation{Center for Nonlinear Studies (CNLS)/T-4, Theoretical Division,
Los Alamos National Laboratory, Los Alamos, NM 87545}
\affiliation{Department of Chemistry and Biochemistry, MSC 3C, New
Mexico State University, Las Cruces, NM 88003}
\author{Michael Thoss}
\affiliation{Institut f\"ur Theoretische Physik and Interdisziplin\"ares
Zentrum f\"ur Molekulare Materialien, Friedrich-Alexander-Universit\"at
Erlangen-N\"urnberg,\\ 
Staudtstr.\,7/B2, D-91058 Erlangen, Germany}
\author{Haobin Wang\footnote{electronic mail: whb@intrepid.nmsu.edu}}
\affiliation{Department of Chemistry and Biochemistry, MSC 3C, New
Mexico State University, Las Cruces, NM 88003}

%\date{\today}

\begin{abstract}

An analog of the Meir-Wingreen formula for the steady-state heat current through
a model molecular junction is derived. The expression relates the heat current to
correlation functions of operators acting only on the degrees of freedom of
the molecular junction. As a result, the macroscopic heat reservoirs are not
treated explicitly. This allows one to exploit methods based on a reduced
description of the dynamics of a relatively small part of the overall system to
evaluate the heat current through a molecular junction.
The derived expression is applied to calculate the steady-state heat current in the weak 
coupling limit, where Redfield theory is used to describe the reduced dynamics of the 
molecular junction. The results are compared with those of previously developed 
approximate and numerically exact methods.

\end{abstract}

\maketitle

\section{Introduction}

Heat transport in nanoscale molecular junctions, i.e., in molecules that
interconnect metal or semiconductor electrodes, is a process that is crucial 
for the stability of the junction and thus for potential molecular electronic 
devices.\cite{Montgomery2002-5377,Chen2003-1691,Segal2003-6840,
Huang2006-1240,Pecchia2007-035401,Ioffe2008-727,Hartle2009-146801}
It has been demonstrated experimentally that the localized Joule heating 
may induce a substantial temperature increase within a molecule-metal contact due to 
inefficient heat dissipation.\cite{Huang2006-1240} Theoretical studies of heat
transport at the nanoscale, in particular the dependence of heat 
dissipation on various physical parameters of a molecular junction, will thus provide valuable
insight into the transport mechanisms thus facilitating the interpretation of
experimental results and the design
of novel nanoscale electronic devices.

Segal and Nitzan investigated the characteristics of heat transport of a spin-boson
nanojunction model (SBNM), where a two-level system is simultaneously connected to
two heat reservoirs (baths) of different temperatures.\cite{Segal2005-034301,
Segal2005-194704,Segal2006-026109} Based on these studies, they suggested possible
realization of novel nano-devices such as a thermal rectifier\cite{Segal2005-034301,
Segal2005-194704} and a molecular heat pump.\cite{Segal2006-026109}
The methodology used in their work assumes weak coupling between the two-level system
and the reservoirs so that Redfield theory\cite{Redfield1957-19,Redfield1965-1} may
be applied. This assumption may not always hold for a realistic molecular junction
where the energy flow may be enhanced/maximized by tuning certain physical parameters.
To address this problem, we have developed a numerically exact
methodology\cite{Velizhanin2008-325} to study the dynamics of the SBNM.
The methodology is based on the multilayer multiconfiguration time-dependent
Hartree (ML-MCTDH) theory,\cite{Wang2003-1289} which is a non-perturbative and
numerically exact approach. Thus, it can provide accurate results in a broader range
of physical regimes (within the model). For example, our previous study has revealed
a turnover behavior of the heat current with respect to the coupling strength between
the two-level system and the heat baths. As a consequence, the optimization of heat
transport is possible by choosing an appropriate set of physical parameters.

Numerically exact simulations\cite{Velizhanin2008-325} also provide 
benchmark results that can be used to develop more accurate approximate
theories. 
This is the focus of the present paper.
From a broader perspective, the spin-boson nanojunction model is a 
two-bath, nonequilibrium version of the standard spin-boson Hamiltonian for studying
electron transfer reactions.\cite{Leggett1987-1,Weiss1999} To date, the reduced dynamics of the two-level system has been of primary interest
and various approximate approaches have been developed to study this dynamics.
Examples include the noninteracting blip
approximation,\cite{Leggett1987-1,Aslangul1986-1657}  Redfield
theory,\cite{Redfield1957-19,Redfield1965-1} Zusman equation\cite{Zusman1980-295}
and its generalization to the low temperature domain.\cite{Ankerhold2004-1436}
It is desirable to apply these methods directly to the study of heat transport
in the spin-boson nanojunction model. However, this is not straightforward since the
operator for heat current involves not only the degrees of freedom of the two-level
system but also those of the heat baths.

On the other hand, the relation between the reduced dynamics of an open system and 
the current flowing through it was found in the theory of electron transport in
mesoscopic systems by Meir and Wingreen.\cite{Meir1992-2512}
They derived a Landauer-type expression that relates the electrical current
to certain correlation functions of the small mesoscopic system.
The goal of this paper is to derive a similar expression for the heat current through
a two-level system, driven out of equilibrium by two heat baths.
Based on the thus obtained expression, the approximate methods mentioned above 
can be used to study heat transport properties of the SBNM. 

The remainder of the paper is organized as follows. The model Hamiltonian and the observable
of interest are described in Sec.~\ref{SBNM_sec}. Section \ref{MDyson_sec} describes 
the theoretical techniques that allow us to express the heat current through
correlation functions of the two-level system. The final Meir-Wingreen type expression
for the steady-state heat current is obtained in Section~\ref{SS_sec}.
As a demonstration, we apply the developed theory in 
Sec.~\ref{Redf_sec} to evaluate the heat current in the limit of weak coupling 
between the two-level system and the baths. In Sec.~\ref{NumRes_sec} this heat current 
is compared with the results obtained using the Segal-Nitzan approach and 
the numerically exact simulations. Sec~\ref{Concls_sec} concludes.

\section{Theory}

\subsection{Spin-Boson nanojunction model}\label{SBNM_sec}

\begin{figure}[!ht]
\begin{center}

\vspace{0.3in}

\epsfig{file=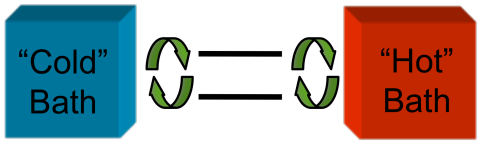,width=2.5in}

\vspace{-0.2in}

\end{center}
\caption{\label{SBNM_Fig} Spin-Boson nanojunction model.}
\vspace{0.2in}
\end{figure}

The spin-boson nanojunction model (SBNM) considered in this work
\cite{Segal2005-034301,Segal2005-194704,Segal2006-026109,Velizhanin2008-325}
consists of two heat reservoirs (baths) interconnected by a bridge, which is
represented by a two-level system (TLS)
(see Fig.~\ref{SBNM_Fig}).
%------------------------------------------------------------------------
%------------------------------------------------------------------------
The heat baths at different temperatures drive the TLS out of equilibrium and the dependence
of the heat current through the TLS on various parameters of the model can be studied.
The SBNM Hamiltonian reads as
\begin{equation}\label{HamTot}
H=H_B+H_S+H_{SB},
\end{equation}
where $H_B$ describes the two harmonic baths, ``cold'' ($C$) and ``hot'' ($H$)
(atomic units are assumed throughout the paper)
\begin{gather}
H_B=H_C+H_H,\nonumber \\
H_K=\sum_{m\in K} \omega_m a^\dagger_m a_m;~~K=C,H,
\end{gather}
where $a^\dagger_m$ ($a_m$) are bosonic creation (annihilation) operators.
The bridge Hamiltonian $H_{S}$ in the second quantization reads
\begin{equation}\label{TLS}
H_{S}=\sum_{i=1,2} E_i n_{ii};~~E_2-E_1=\epsilon>0,
\end{equation}
where $n_{ij}\equiv c^\dagger_i c_j$ and $c^\dagger_i$ ($c_j$) is
a fermionic creation (annihilation) operator. The energy spacing between
the two levels of the bridge is denoted by $\epsilon$. The coupling between
the bath and the TLS is given by
\begin{gather}
H_{SB}=H_{SC}+H_{SH},\nonumber \\
H_{SK}=\sum_{m\in K} V_m (a^\dagger_m+a_m)[n_{21}+n_{12}];~~K=C,H.
\label{HamCoupl}
\end{gather}

The dynamics of the TLS is restricted to
the single-particle space spanned by the basis functions $|1\rangle=c^\dagger_1|{\rm vac}\rangle$ and
$|2\rangle=c^\dagger_2|{\rm vac}\rangle$, where $|{\rm vac}\rangle$ denotes the fermionic vacuum.
This restriction guarantees that the TLS representation by Hamiltonian~(\ref{TLS})
is equivalent to the more commonly used form,\cite{Leggett1987-1,Weiss1999}
$H_S=\frac{\epsilon}{2}(|2\rangle\langle 2|-|1\rangle\langle 1|)$.
On the other hand, the fermionic representation introduced here is more convenient
for the diagrammatic technique exploited in Sec.~\ref{MDyson_sec}.

\begin{figure}[!ht]
\begin{center}

\vspace{0.3in}

\epsfig{file=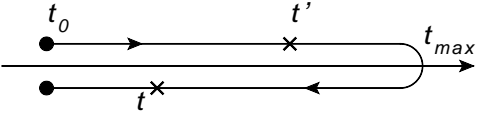,width=2.5in}

\vspace{-0.2in}

\end{center}
\caption{\label{Contour} The Schwinger-Keldysh contour.}
\vspace{0.2in}
\end{figure}

Starting with the Heisenberg operator of the heat current from the TLS to the cold bath
\begin{equation}
I_C(t)\equiv i[H,H_C](t)=
i\left[H,\sum_{m\in C} \omega_m a^\dagger_m a_m\right](t),
\end{equation}
straightforward algebra gives the expectation value of the heat current
\begin{equation}\label{Hflux}
\langle I_C(t)\rangle=-2{\rm Im}\left[\sum_{i,j;~m\in C}
\omega_m V^{ij}_m M^{>}_{m,ij}(t,t)\right ].
\end{equation}
In this expression we introduce the notation
\begin{equation}
V^{ij}_m=(1-\delta_{ij})V_m,
\end{equation}
and the nonequilibrium Green's function (NEGF) $M_{m,ij}(t,t^\prime)$
defined on the Schwinger-Keldysh contour\cite{Schwinger1961-407,Keldysh1965-1018}
(depicted in Fig.~\ref{Contour}) 
\begin{subequations}\label{COGF}
\begin{equation}\label{Mcorr}
M_{m,ij}(t,t^\prime)=\langle T_c[a_m(t)
n_{ij}(t^\prime)]\rangle
\end{equation}
\begin{equation}
M_{m,ij}(t,t^\prime)=\left \{\begin{array}{ll}
M^>_{m,ij}(t,t^\prime);~~t>_c t', \\
M^<_{m,ij}(t,t^\prime);~~t<_c t', \\
\end{array}\right.
\end{equation}
\end{subequations}
where $t>_c t^\prime$ means that $t$ is "later" on the contour than $t^\prime$
as is illustrated in Fig.~\ref{Contour}.
%--------------------------------------------------------------------
%--------------------------------------------------------------------
The contour-ordering operator $T_c$ moves operators with later (in the contour sense)
time-arguments to the left. Thus, in accordance with Eq.~(\ref{COGF}), the greater and
the lesser Green's functions
are defined as $M^>_{m,ij}(t,t')=\langle a_m(t) n_{ij}(t')\rangle$ and 
$M^<_{m,ij}(t,t')=\langle n_{ij}(t') a_m(t)\rangle$, respectively.
The choice of $t_{max}$, the contour's turning point, is arbitrary
as long as $t_{max}>t,t'$.
The quantum mechanical average with respect to the initial state in 
Eq.~(\ref{Mcorr}) is defined as $\langle ...\rangle={\rm Tr}[...~\rho(t_0)]$,
where the initial state density operator $\rho(t_0)$ is taken in the product form
\begin{subequations}
\begin{equation}
\rho(t_0)=\rho_B\times\rho_S,
\end{equation}
\begin{equation}\label{rhoB}
\rho_B=\frac{e^{-\beta_C H_{C}}e^{-\beta_H H_{H}}}{{\rm Tr}\left \{
e^{-\beta_C H_{C}}e^{-\beta_H H_{H}}\right\} }.
\end{equation}
\end{subequations}
In the expression above $\beta_C=1/k_B T_C$, $\beta_H=1/k_B T_H$, and $k_B$
is the Boltzmann constant. The initial density operator
of the bridge $\rho_S$ is arbitrary as long as one is only interested in the
steady state.

\subsection{Dyson equation for $M_{m,ij}(t,t^\prime)$}\label{MDyson_sec}

Inspired by the pioneering work of Meir and Wingreen,\cite{Meir1992-2512}
we proceed to express $M_{m,ij}(t,t^\prime)$ in terms of correlation
functions of the TLS operators only. Our derivation closely follows
the procedure used by Gaudin in his proof of the generalized Wick's
theorem.\cite{Gaudin1960-89,Fetter1971} First, we rewrite $M_{m,ij}(t,t^\prime)$
in the interaction representation
\begin{equation}
M_{m,ij}(t,t^\prime)={\rm Tr}\left \{S_c[ \hat{a}_m(t)
\hat{n}_{ij}(t^\prime)]\rho(t_0)\right \}.
\end{equation}
Operators in the interaction representation are denoted by a hat, i.e.,
$\hat{a}_m(\tau)=e^{i(H_S+H_B)\tau}a_m e^{-i(H_S+H_B)\tau}$,
and $S_c$ is 
\begin{align}
S_c&=T_c \left [ 1+(-i)\int_c d\tau\:\hat{H}_{SB}(\tau)\right.
\nonumber \\
&\left.+\frac{(-i)^2}{2}\int_c d\tau\int_c
d\tau^\prime\: \hat{H}_{SB}(\tau)\hat{H}_{SB}(\tau^\prime)+
\dots \right ].
\label{SProp_SKC}
\end{align}
The above expansion of $S_c$ in powers of $\hat{H}_{SB}(\tau)$
yields the perturbation expansion for $M_{m,ij}(t,t^\prime)$
\begin{equation}\label{Mterms}
M_{m,ij}(t,t^\prime)=M^{(0)}_{m,ij}(t,t^\prime)+M^{(1)}_{m,ij}(t,t^\prime)+\dotsb,
\end{equation}
where
\begin{gather}
M^{(0)}_{m,ij}(t,t^\prime)=
{\rm Tr}\left\{T_c[\hat{a}_m(t)\hat{n}_{ij}(t^\prime)]\rho(t_0)\right\}=0, \nonumber \\
M^{(1)}_{m,ij}(t,t^\prime)=-i\int_c d\tau \; 
{\rm Tr}\left\{T_c[\hat{H}_{SB}(\tau)\hat{a}_m(t)\hat{n}_{ij}(t^\prime)]
\rho(t_0)\right\},
\nonumber \\
\dotso
\end{gather}
The successive commutation of $a_m$ through all operators $\hat{H}_{SB}$ and $\rho(t_0)$,
with the use of the following identity \cite{Gaudin1960-89,Fetter1971}
\begin{equation}
\exp(-\beta\omega_m a^\dagger_m a_m a_m)=
a_m \exp(-\beta\omega_m a^\dagger_m a_m)\exp(-\beta\omega_m),
\end{equation}
and the cyclic permutation within the trace, leads to
\begin{gather}
M^{(1)}_{m,ij}(t,t^\prime)=-i\sum_{kl}V_m^{kl}\int_c d\tau\; 
{\rm Tr}\left\{ T_c[\hat{a}_m(t)\hat{a}^\dagger_m(\tau)]\rho(t_0)\right\}
\nonumber \\ \times
{\rm Tr}\left\{T_c[\hat{n}_{kl}(\tau)\hat{n}_{ij}(t^\prime)]\rho(t_0)\right\},\nonumber \\
M^{(2)}_{m,ij}(t,t^\prime)=(-i)^2\sum_{kl}V_m^{kl}\int_c d\tau\; 
{\rm Tr}\left\{T_c[\hat{a}_m(t)\hat{a}^\dagger_m(\tau)]\rho(t_0)\right\}
\nonumber \\ \times
\int_c d\tau^\prime\;{\rm Tr}\left\{T_c[\hat{H}_{SB}(\tau^\prime)\hat{n}_{kl}
(\tau)\hat{n}_{ij}(t^\prime)]\rho(t_0)\right\},\nonumber \\ 
\dotso \label{Mexpan}
\end{gather}
By continuing the series in Eq.~(\ref{Mexpan}) one notices that
\begin{equation}\label{MKexpan}
M^{(q)}_{m,ij}(t,t^\prime)=\sum_{kl} V_m^{kl}\int_c
d\tau\: D_m(t,\tau)K^{(q-1)}_{kl,ij}(\tau,t^\prime),
\end{equation}
where
\begin{equation}\label{BathGF}
D_m(t,\tau)=-i{\rm Tr}\left\{T_c[\hat{a}_m(t)\hat{a}^\dagger_m(\tau)]
\rho(t_0)\right\}
\end{equation}
is the contour-ordered Green's function of the non--interacting bath and
$K^{(q)}_{kl,ij}(\tau,t^\prime)$ arises from the perturbation expansion
of the Green's function
\begin{equation}\label{Kexpan}
K_{kl,ij}(\tau,t^\prime)=\sum^{\infty}_{q=0}K^{(q)}_{kl,ij}(\tau,t^\prime)
={\rm Tr}\left\{T_c[n_{kl}(\tau)n_{ij}(t^\prime)]\rho(t_0)\right\}.
\end{equation}
Finally, substitution of Eqs.~(\ref{MKexpan}) and (\ref{Kexpan}) into
Eq.~(\ref{Mterms}) results in
\begin{equation}\label{MDyson}
M_{m,ij}(t,t^\prime)=\sum_{kl}V_m^{kl}
\int_c d\tau \; D_m(t,\tau)K_{kl,ij}(\tau,t^\prime),
\end{equation}
which expresses $M_{m,ij}(t,t^\prime)$ in terms of correlation functions
of TLS operators and non-interacting bath operators separately.
It is noted that Eq.~(\ref{MDyson}) can also be obtained using the equation-of-motion
technique.\cite{Jauho1994-5528,Haug1996,Bruus2004}

A Dyson equation similar to Eq.~(\ref{MDyson}) was recently obtained in the context of
charge transport through single molecules, where molecular many-body states were
employed to describe the central molecular part.\cite{Esposito2009-205303}
This similarity is due to the formal resemblance of the Hamiltonian describing 
the bath-TLS coupling, Eq.~(\ref{HamCoupl}) in the present work, and that of
molecule-contact coupling, Eq. (8) in Ref.~\onlinecite{Esposito2009-205303}. 

\subsection{Steady-state heat current}\label{SS_sec}

Since only the greater NEGF $M^>_{m,ij}(t,t^\prime)$ is needed to evaluate the
heat current in Eq.~(\ref{Hflux}), we can immediately assign 
$t$ and $t^\prime$ to the lower and the upper Schwinger-Keldysh
contour branches, respectively. Omitting straightforward manipulations 
with Eq.~(\ref{MDyson}) based on Langreth's contour
deformation rules,\cite{Haug1996} the final result
for $M^>_{m,ij}(t,t^\prime)$, which contains only
traditional real-time correlation functions and the time integration along the
real axis, is obtained
\begin{align}
M^>_{m,ij}(t,t^\prime)&=\sum_{kl}V_m^{kl}\int^\infty_{t_0}d\tau\left[
D^r_m(t,\tau)K^>_{kl,ij}(\tau,t^\prime)\right.
\nonumber \\
&\left.+D^>_m(t,\tau)K^a_{kl,ij}(\tau,t^\prime)\right ].
\label{Mgreater}
\end{align}
Here, correlation functions $D(t,\tau)$ and $K(\tau,t')$ are given in
Eqs.~(\ref{BathGF}) and (\ref{Kexpan}), respectively, and for an arbitrary contour-ordered
Green's function $G(t,t^\prime)$ we define
\begin{align}
G^r(t,t')&\equiv\theta(t-t')[G^>(t,t')-G^<(t,t')],
\nonumber \\
G^a(t,t')&\equiv\theta(t'-t)[G^<(t,t')-G^>(t,t')],
\end{align}
where $\theta(t-t')=1$ if $t> t'$ and zero otherwise.

In the steady-state regime ($t_0\rightarrow -\infty$) two-time 
correlation functions become stationary, i.e.,
$M^>_{m,ij}(t,t^\prime)=M^>_{m,ij}(t-t^\prime)$.
Combining Eq.~(\ref{Hflux}) with Eq.~(\ref{Mgreater}) and setting $t=0$
for simplicity we obtain the expression for the heat current
\begin{gather}
\langle I_C \rangle = -2{\rm Im} \sum_{i,j;k,l}\sum_{m\in C} \omega_m V_m^{ij}V_m^{kl}
\nonumber \\
\times \int^{+\infty}_{-\infty}d\tau
\left (D^r_m(-\tau)K^>_{kl,ij}(\tau)+D^>_m(-\tau)K^a_{kl,ij}(\tau)\right ).
\end{gather}
The Green's functions of the cold non-interacting bath can be easily evaluated
as \cite{Haug1996,Mahan2000}
\begin{align}
D^r_m(-\tau)&=-i\theta(-\tau)e^{i\omega_m \tau}, \nonumber \\
D^>_m(-\tau)&=-i(1+\eta^C_m)e^{i\omega_m \tau},
\end{align}
where $\eta^C_m=\eta^C(\omega_m)=[e^{\beta_C \omega_m}-1]^{-1}$ is the Bose-Einstein
distribution.
Combining the last two equations and noting that the time
integration corresponds to a Fourier transform we finally obtain
\begin{align}
\langle I_C \rangle&=\sum_{i,j;k,l}\sum_{m\in C} \omega_m V_m^{ij}V_m^{kl}
\left [(1+\eta^C_m)K^<_{kl,ij}(\omega_m)\right.
\nonumber \\
&\left.-\eta^C_m K^>_{kl,ij}(\omega_m)\right ],
\label{SSflux}
\end{align}
where
\begin{equation}
K^{>(<)}_{kl,ij}(\omega)=\int^{+\infty}_{-\infty} d\tau\:
K^{>(<)}_{kl,ij}(\tau)e^{i\omega\tau}.
\end{equation}
The expression for the heat current to the other (hot) bath is obtained 
by a straightforward substitution of ``C'' with ``H'' in Eq.~(\ref{SSflux})

Eq.~(\ref{SSflux}) {\it exactly} relates
the heat current in the spin-boson nanojunction model to correlation functions
of the bridge that can be evaluated using methods developed to describe
reduced dynamics of a TLS. In fact, Eq.~(\ref{SSflux}) is also formally valid for
a multi-level bridge. In this case the indices $i,j,k,l$ run through all multiple bridge
states and the expression includes the corresponding coupling constants $V^{ij}_m$ and correlation
functions $K^{>(<)}_{kl,ij}(\omega)$. A similar expression was recently derived for
{\em photonic} heat current through an arbitrary (nonlinear) circuit element coupled
to two dissipative reservoirs at finite
temperatures.\cite{Ojanen2008-155902,Ruokola2009-144306}

The next subsection is dedicated to an approximate evaluation of the TLS
correlation functions, and hence the SBNM heat current, in the limit of weak
TLS-bath coupling.

\subsection{Treatment of the correlation function within the Redfield
approximation}\label{Redf_sec}

Redfield theory in the form of kinetic equations for the TLS populations has been
previously employed by Segal and Nitzan to study the heat-conducting
properties of the SBNM.\cite{Segal2005-034301,Segal2005-194704,Segal2006-026109}
Within the Redfield approximation it is possible to obtain simple
and physically transparent analytical results. However, this
approach is accurate only at very weak TLS-bath couplings, which restricts its applicability.
In particular, the non-monotonic dependence of the heat current
on the coupling strength (see the next section for details) can not be described within
this approach.\cite{Velizhanin2008-325}

In the remainder of the paper we will use the theory developed
in the previous subsections to derive an improved description of heat
transport that still retains the analytical nature of the method. Specifically, we will evaluate the
TLS correlation functions $K^{>(<)}_{kl,ij}(\omega)$ within the Redfield approximation.
The thus evaluated correlation functions will then be used to calculate the heat current by
means of Eq.~(\ref{SSflux}). As will be demonstrated in
Sec.~\ref{NumRes_sec}, this approach results in a notable improvement over the
original Redfield-type theory by Segal and Nitzan. 

In the following discussions it is convenient to split $K^{>(<)}_{kl,ij}(t,t^\prime)$ into
the ``retarded'' ($+$) and ``advanced'' ($-$) part
%-----------------------------------------------------------------------------------
\begin{equation}
K^>_{kl,ij}(t,t^\prime)=K^+_{kl,ij}(t,t^\prime)+K^-_{kl,ij}(t,t^\prime),
\end{equation}
where
\begin{subequations}\label{Kterms}
\begin{equation}
K^+_{kl,ij}(t,t^\prime)=\theta(t-t^\prime){\rm Tr}\left\{e^{iH(t-t^\prime)}
n_{kl}e^{-iH(t-t^\prime)}n_{ij}\rho(t^\prime)\right \},
\end{equation}
\begin{equation}
K^-_{kl,ij}(t,t^\prime)=\theta(t^\prime-t){\rm Tr}\left\{e^{iH(t^\prime-t)}
n_{ij}e^{-iH(t^\prime-t)}\rho(t)n_{kl}\right\}.
\end{equation}
\end{subequations}
%---------------------------------------------------------------------------
Lesser correlation function $K^<_{kl,ij}(t,t^\prime)$ can always
be obtained using the property $K^<_{kl,ij}(t,t^\prime)=K^>_{ij,kl}(t^\prime,t)$.
Furthermore, the retarded and the advanced parts are related through
$K^-_{ji,lk}(t',t)=\left [K^+_{kl,ij}(t,t')\right ]^*$,
so we focus on the retarded correlation function and obtain all other functions accordingly.
The steady-state regime implies that
both $t-t_0$ and $t^\prime-t_0$
significantly exceed the characteristic time of the transient dynamics.
This, together with the Redfield approximation,\cite{Redfield1957-19,Redfield1965-1}
yields the steady-state density operators in Eq.~(\ref{Kterms})
\begin{equation}\label{ssrho}
\rho(t)=\rho(t^\prime)=\left (|1\rangle\langle 1|P^S_1+
|2\rangle\langle 2|P^S_2\right)\rho_B,
\end{equation}
where $P^S_1$, $P^S_2$ are the stationary populations of
the two levels and
$\rho_B$ is defined in Eq~(\ref{rhoB}).
Since the density operators $\rho(t)$ and $\rho(t')$ in Eq.~(\ref{ssrho}) are effectively
time-independent, the correlation functions in Eq.~(\ref{Kterms})
depend only on $\tau=t-t^\prime$ as expected at the steady state.

The correlation functions $K^{+}_{ij,kl}$ have a clear physical
meaning. For example $K^{+}_{12,12}(\tau)$ describes a process where the system,
being in the stationary state, undergoes a coherence-introducing transition
$|2\rangle\langle 2|\rightarrow|1\rangle\langle 2|$, then evolves for time $\tau$,
and finally the amplitude of the coherence $|2\rangle\langle 1|$ is measured and
multiplied by the steady-state population of state $|2\rangle\langle 2|$,
i.e., $P^S_2$. Therefore, within the Redfield approximation, this
correlation function can be calculated as
\begin{equation}
K^{+}_{12,12}(\tau)=P^S_2\langle 2|\rho_S(\tau)|1\rangle,
\end{equation}
where $\rho_S(\tau)$ is the solution of the Redfield master equation
\begin{equation}\label{RME}
\frac{d\rho_{ij}(\tau)}{d\tau}=-iE_{ij}\rho_{ij}(\tau)
+\sum_{kl}R_{ijkl}\rho_{kl}(\tau)
\end{equation}
subject to the initial condition $\rho_S(0)=|1\rangle\langle 2|$.
In this master equation, $\rho_{ij}=\langle i|\rho(\tau)|j\rangle$,
$E_{ij}=E_i-E_j$ and the Redfield tensor $R_{ijkl}$ is given by
\begin{equation}
R_{ijkl}=\Gamma^+_{ljik}+\Gamma^-_{ljik}-\delta_{jl}\sum_{\alpha}
\Gamma^+_{i\alpha\alpha k}-\delta_{ik}\sum_{\alpha}\Gamma^-_{l\alpha\alpha j},
\end{equation}
with
\begin{align}
\Gamma^+_{ljik}&=\int^\infty_0 ds \: {\rm Tr}_B\left\{\langle l|e^{-iH_B s}
H_{SB}e^{iH_B s}|j\rangle\langle i|H_{SB}|k\rangle\right\}
\nonumber \\
&\times e^{-iE_{ik} s},
\nonumber \\
\Gamma^-_{ljik}&=\int^\infty_0 ds \: {\rm Tr}_B\left\{\langle l|H_{SB}|j\rangle
\langle i|e^{-iH_B s}H_{SB}e^{iH_B s}|k\rangle\right\}
\nonumber \\
&\times e^{-iE_{lj} s}.
\end{align}
All the other retarded correlation functions can be obtained in a similar fashion.

Instead of solving Eq.~(\ref{RME}) for each correlation
function independently, it is more practical to reformulate the problem
as an equation of motion for the correlation functions themselves.
Straightforward but somewhat tedious algebraic manipulations,
including the evaluation of various components 
$\Gamma^{+(-)}_{ljik}$ of the Redfield
tensor, allow one to obtain this equation of motion 
in matrix form with the initial conditions included
\begin{align}
\frac{d}{d \tau}{\bf K}^+(\tau)&=\begin{pmatrix}
i\epsilon^\prime-\gamma~~ & i\Delta\epsilon+\gamma \\
-i\Delta\epsilon+\gamma~~ & -i\epsilon^\prime-\gamma 
\end{pmatrix}{\bf K}^+(\tau)
\nonumber \\
&+\delta(\tau)\begin{pmatrix}
P^S_2 & 0 \\
0 & P^S_1 \end{pmatrix},
\label{RedODE}
\end{align}
where ${\bf K}^+$ is $2\times 2$ matrix
\begin{equation}
{\bf K}^+(\tau)=\begin{pmatrix}
K^+_{21,12}(\tau)~ & K^+_{21,21}(\tau) \\
K^+_{12,12}(\tau)~ & K^+_{12,21}(\tau) \end{pmatrix}.
\end{equation}
The renormalized energy splitting $\epsilon^\prime$ of the TLS levels is given by
\begin{gather}
\epsilon^\prime=\epsilon+\Delta\epsilon,\nonumber \\
\Delta\epsilon=\frac{1}{\pi}\sum_{K=C,H}\mathcal{P}\int^\infty_0
d\omega\left [1+2\eta_K(\omega)\right ]
\nonumber \\
\times J_K(\omega)\left\{
\frac{1}{\epsilon+\omega}+\frac{1}{\epsilon-\omega}\right\},
\label{renorme}
\end{gather}
where $\mathcal{P}$ stands for the Cauchy principal value.
The spectral density
\begin{equation}
J_K(\omega)=\pi\sum_{m\in K}V^2_m\delta(\omega-\omega_m);~~K=C,H,
\end{equation}
is introduced here to change the summation over the bath modes to
an integration over the frequency in Eq.~(\ref{renorme}).
In this work we employ a spectral density of Ohmic form with
exponential cutoff
\begin{equation}\label{SpecDens}
J_K(\omega)=\lambda_K\frac{\pi \omega}{4\omega_K}e^{-\omega/\omega_K},
\end{equation}
where $\lambda_K$ and $\omega_K$ are the TLS-bath coupling strength
and the characteristic frequency of the $K^{\rm th}$ bath, respectively.
The stationary dephasing rate is related to the spectral density as
\begin{equation}
\gamma=\sum_{K=C,H}\left[1+2\eta_K(\epsilon)\right.]J_K(\epsilon)
\end{equation}

The system of linear ordinary differential equations in Eq.~(\ref{RedODE})
can be easily solved by converting it to a system of linear algebraic equations
employing a Fourier transform
\begin{equation}\label{Ksolved}
{\bf K}^+(\omega)=\frac{1}{D(\omega)}\begin{bmatrix}
(\gamma+i\epsilon^\prime-i\omega)P^S_2, &
(\gamma+i\Delta\epsilon)P^S_1 \\
(\gamma-i\Delta\epsilon)P^S_2, &
(\gamma-i\epsilon^\prime-i\omega)P^S_1
\end{bmatrix},
\end{equation}
where
$D(\omega)=(\epsilon^\prime)^2-(\Delta\epsilon)^2-2i\omega\gamma
-\omega^2$.
The inverse Fourier transform is not needed since these correlation functions
in frequency domain are exactly what is required to evaluate the heat current in
Eq.~(\ref{SSflux}). The expression for ${\bf K}^+$ is the only one explicitly
needed since ${\bf K}^-=\left({\bf K}^+\right)^\dagger$ 
and the greater and lesser correlation functions are given by
\begin{gather}
K^>_{kl,ij}(\omega)=K^+_{kl,ij}(\omega)+K^-_{kl,ij}(\omega), \nonumber \\
K^<_{kl,ij}(\omega)=K^>_{ij,kl}(-\omega). \label{Krels}
\end{gather}
Finally, Eq.~(\ref{SSflux}) can be rewritten with the use of 
Eqs.~(\ref{Ksolved}) and (\ref{Krels})
\begin{equation}\label{Flux_NEGF_Redf}
\langle I_C\rangle=\Gamma^C_d P^S_2-\Gamma^C_u P^S_1,
\end{equation}
where heat transfer rates are
\begin{subequations}\label{HeatTrRates}
\begin{equation}
\Gamma^C_d=\frac{2}{\pi}\int^\infty_0 d\omega\: J_C(\omega)\omega\times
{\rm Re}\left\{\frac{2\gamma-i\epsilon(1+2\eta_C(\omega))-i\omega}
{D(\omega)}\right \},
\end{equation}
\begin{equation}
\Gamma^C_u=-\frac{2}{\pi}\int^\infty_0 d\omega\: J_C(\omega)\omega\times
{\rm Re}\left\{\frac{2\gamma+i\epsilon(1+2\eta_C(\omega))-i\omega}
{D(\omega)}\right \}.
\end{equation}
\end{subequations}
The heat transfer rates for the hot bath can be found from the expressions above
by formal substitution of ``C'' with ``H''.

In a weak coupling regime the heat transfer rates reduce to
\begin{equation}\label{HeatTrRates_Redf}
\Gamma^C_d=2\epsilon J_C(\epsilon)\left[1+\eta_C(\epsilon)\right],~
\Gamma^C_u=2\epsilon J_C(\epsilon)\eta_C(\epsilon),
\end{equation}
which are identical to the results of Ref.~\onlinecite{Segal2005-034301}. 
Another way to obtain the rates in Eq.~(\ref{HeatTrRates_Redf})
is to eliminate the dephasing (and
energy renormalization) by setting $\gamma$ and $\Delta\epsilon$
in Eq.~(\ref{RedODE}) to zero.
Hence, the previous expression\cite{Segal2005-034301} for the heat transfer rate is recovered 
when coherence, described by correlation functions $K(\tau)$, persists indefinitely. Once 
realistic dephasing
and energy renormalization are added, Eq.~(\ref{HeatTrRates}) for the heat current
is obtained. From this we expect that Eq.~(\ref{HeatTrRates})
can provide a more accurate description of the SBNM heat current than
Eq.~(\ref{HeatTrRates_Redf})
except for the very weak coupling regime where they agree with each other.
In what follows, we will refer to the expression for the heat current obtained
in this work [Eqs.~(\ref{Flux_NEGF_Redf}), (\ref{HeatTrRates})] as NEGF-Redfield approach
to distinguish it from the Redfield approximation used previously.\cite{Segal2005-034301}

Finally, it is worthwhile to discuss the energy conservation in the NEGF-Redfield approach.
Conservation laws are not always automatically satisfied in approximate theories.
Thus, often special care must be taken to guarantee conservation of, e.g.,
number of particles, energy or momentum.
For example, in the application of many-body Green's functions it has been demonstrated that
self-consistency must be incorporated into the Dyson equation in order to preserve
conservation laws.\cite{Baym1961-287,Baym1962-1391} As will be seen later,
the NEGF-Redfield approach developed in this subsection does not conserve energy, i.e., 
$\langle I_C\rangle+\langle I_H\rangle\neq 0$, if the TLS steady-state populations
are determined as the stationary solution of Eq.~(\ref{RME}). 
In fact, numerical tests show that with so obtained steady-state
populations the heat current in Eq.~(\ref{Flux_NEGF_Redf}) does not necessarily
vanish at equilibrium, i.e.,when the both baths have identical temperature.
To correct for this, we introduce self-consistency by explicitly requiring
the energy conservation
\begin{equation}\label{SelfCons}
\langle I_C\rangle+\langle I_H\rangle=0,
\end{equation}
The steady-state populations are then determined from this constraint.
Specifically, once the heat transfer rates, Eq.~(\ref{HeatTrRates}), are evaluated
for both baths, Eq.~(\ref{SelfCons}) becomes a linear equation for
the steady-state populations of the two levels. This linear equation is solved straightforwardly
taking into account the identity $P^S_1+P^S_2=1$.
It is noted that the so obtained steady-state populations are identical to those
obtained by assuming a kinetic equation for the TLS populations with
the ``cold'' and ``hot'' rates given by Eq.~(\ref{HeatTrRates}).
The latter procedure can be straightforwardly extended to multi-level bridges.

It will be demonstrated in the next section that the introduction of self-consistency 
performs surprisingly well and drastically improves the accuracy of the NEGF-Redfield method.

\section{Numerical Results and Discussion}\label{NumRes_sec}

\begin{figure}[!ht]
\vspace{0.3in}
\begin{center}

\epsfig{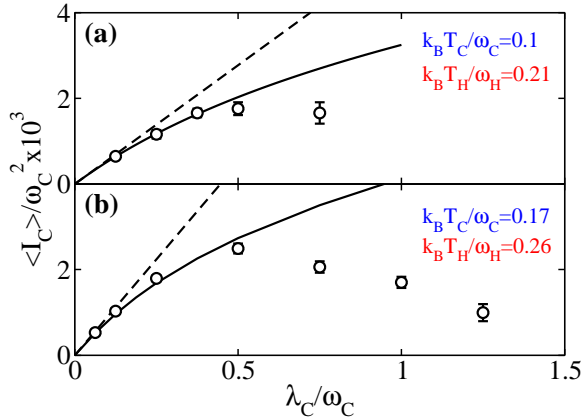}

\vspace{-0.2in}

\end{center}
\caption{\label{HT_TurnOver_1} Dependence of the steady-state
heat current $I_C$ on TLS-bath coupling strength $\lambda_C$.
Panels (a) and (b) correspond to different temperatures of baths.
The spectral densities for the two baths are equal 
($\lambda_C=\lambda_H$, $\omega_C=\omega_H$). The TLS energy spacing
is $\epsilon/\omega_C=0.75$. The numerical results have been obtained with
the numerically exact ML-MCTDH methodology (circles), the Segal-Nitzan theory (dashed lines),
and the NEGF-Redfield approach developed in this work (full lines). Error bars for some numerically
exact points are smaller than the size of circles, and, therefore, not shown in the figure.}
\vspace{0.2in}
\end{figure}

The dependence of the steady-state heat current $\langle I_C\rangle$
on the TLS-bath coupling strength $\lambda_C$
is shown in Figure~\ref{HT_TurnOver_1}.
%--------------------------------------------------------------------
%--------------------------------------------------------------------
The bath parameters in Eq.~(\ref{SpecDens}), i.e., the TLS-bath coupling strength
and the characteristic frequency, are chosen to be identical for the two baths, i.e., $\lambda_C=\lambda_H$
and $\omega_C=\omega_H$, respectively. The bath temperatures are defined by $k_B T_C/\omega_C=0.1$,
$k_B T_H/\omega_H=0.21$ for the cold and the hot bath in panel (a) and $k_B T_C/\omega_C=0.17$,
$k_B T_H/\omega_H=0.26$ for the cold and the hot bath in panel (b), respectively.
The ratio of the TLS energy spacing to the characteristic frequency of the bath is
taken as $\epsilon/\omega_C=0.75$ in both panels.

The numerical results of the NEGF-Redfield and Segal-Nitzan methods are obtained from 
Eq.~(\ref{HeatTrRates}) and Eq.~(\ref{HeatTrRates_Redf}), respectively.
The numerically exact results are obtained
by means of the Multilayer Multiconfiguration Time-Dependent Hartree (ML-MCTDH)
approach.\cite{Wang2003-1289} The details on the application of the ML-MCTDH method
to the study of heat transport in the SBNM are described elsewhere.\cite{Velizhanin2008-325}
For the parameters considered here both the NEGF-Redfield and Segal-Nitzan theories,
depicted by the dashed and full black lines, perform well if $\lambda_C$
is sufficiently small.
However, when $\lambda_C/\omega_C$ is larger than
$\sim 0.5$, both approximate theories cease to be valid and the results
deviate significantly from the numerically
exact simulation. This failure is expected considering
the weak-coupling nature of the approximations in both approximate methods.
In the intermediate region
($\lambda_C/\omega_C\approx $ 0.3-0.5) the NEGF-Redfield method gives
significantly better results than the Segal-Nitzan approach, because the former treats
the TLS-bath coupling more accurately as was discussed in the previous section.

At large coupling strengths the numerically exact simulation predicts
a ``turnover'',\cite{Hanggi1990-251,Pollak2000-1,Velizhanin2008-325} i.e.,
the heat current reaches its maximum and starts to decrease with increasing of
the coupling strength. This phenomenon is similar to Kramers' turnover
and has been discussed in more
detail elsewhere.\cite{Velizhanin2008-325} The turnover is not
reproduced (even qualitatively) by the approximate theories although 
the NEGF-Redfield method displays the correct curvature change in the intermediate
coupling regime.

\begin{figure}[!ht]
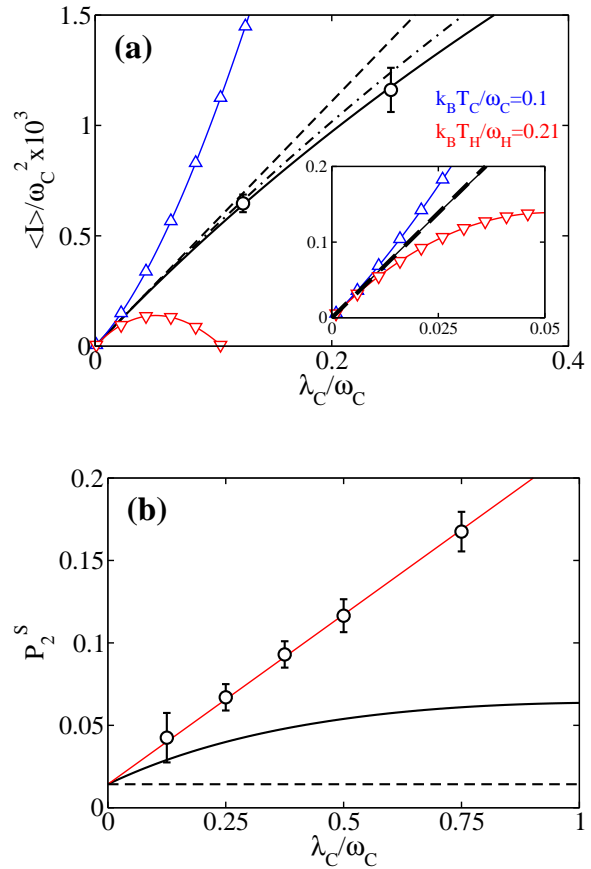

\vspace{0.3in}
\begin{center}

\epsfig{file=Fig4a.eps,width=3.0in}

\vspace{0.31in}

\epsfig{file=Fig4b.eps,width=3.0in}

\vspace{-0.2in}

\end{center}
\caption{\label{HT_TurnOver_2} Dependence of the steady-state
heat current (Panel a) and the steady-state population of the higher TLS level 
(Panel b) on the TLS-bath coupling strength $\lambda_C$.
All parameters are the same and circles, full lines and dashed lines correspond 
to the same theoretical methods as in Fig.~\ref{HT_TurnOver_1}(a).
The up-triangles and down-triangles depict $\langle I_C\rangle$ and
$-\langle I_H\rangle$, respectively, in the case when steady-state populations
are obtained directly from the Redfield master equation, Eq.~(\ref{RME}),
and not from the self-consistency requirement, Eq.~(\ref{SelfCons}).
The average of these two currents, i.e., $(\langle I_C\rangle-\langle I_H\rangle)/2$,
is depicted by the dashed-dotted line. The thin red line in Panel b, which was
obtained by fitting the numerically
exact results, is a guide for the eye.}
\vspace{0.2in}
\end{figure}

The accuracy of the NEGF-Redfield approach deteriorates significantly if 
the self-consistency constraint, Eq.~(\ref{SelfCons}), is not applied and 
the steady-state populations are obtained from regular Redfield theory, i.e.,
from Eq.~(\ref{RME}). The heat currents from the bridge to the cold bath, 
$\langle I_C\rangle$, and from the hot bath to the bridge, $-\langle I_H\rangle$,
evaluated for this choice of the steady-state populations, are depicted by
up-triangles and down-triangles, respectively, in Fig.~\ref{HT_TurnOver_2}(a).
%--------------------------------------------------------------------
%--------------------------------------------------------------------
In the inset, it is seen that at very small coupling strength these two currents
coincide with each other and with the current determined self-consistently.
However, $\langle I_C\rangle$ and $-\langle I_H\rangle$ start to deviate
from each other at $\lambda_C/\omega_C\approx 0.01$, thus demonstrating the
non-conservation of the energy current discussed above. Furthermore, the heat
current from the hot bath to the bridge becomes unphysically negative at
$\lambda_C/\omega_C\approx 0.1$. 

The energy conservation condition in Eq.~(\ref{SelfCons}) suggests another,
more symmetric, definition of 
heat current by averaging $\langle I_C\rangle$ and $-\langle I_H\rangle$,
i.e., $(\langle I_C\rangle-\langle I_H\rangle)/2$.\cite{Velizhanin2008-325}
This averaged current, with $\langle I_C\rangle$ and $-\langle I_H\rangle$
given by up- and down-triangles, respectively, is depicted by the dashed-dotted
line in Fig.~\ref{HT_TurnOver_2}.
This current is seen to agree much better with the numerically exact results than
$\langle I_C\rangle$ and $-\langle I_H\rangle$ separately.
However, this ``averaging'' trick does not resolve the energy non-conservation problem,
and merely conceals it. Furthermore, the NEGF-Redfield approach with self-consistency
introduced (full line) is still in better agreement with the numerically exact calculations. 
Therefore, the results of Fig.~\ref{HT_TurnOver_2}(a) emphasize the necessity of
the energy conservation requirement and support our choice of the steady-state
populations.

This choice is further supported by Fig.~\ref{HT_TurnOver_2}(b), where 
the steady-state population of the higher TLS state, evaluated 
from Eq.~(\ref{SelfCons}) within the NEGF-Redfield theory (full line),
Redfield master equation (dashed line) and the numerically exact
simulation (circles), is depicted. Whereas standard Redfield theory (dashed
line) yields a population $P^S_2$
that is totally independent on the coupling strength, the NEGF-Redfield theory performs better
reproducing, although not quantitatively, the increase of the population with $\lambda_C$.

\begin{figure}[!ht]
\begin{center}

\vspace{0.5in}

\epsfig{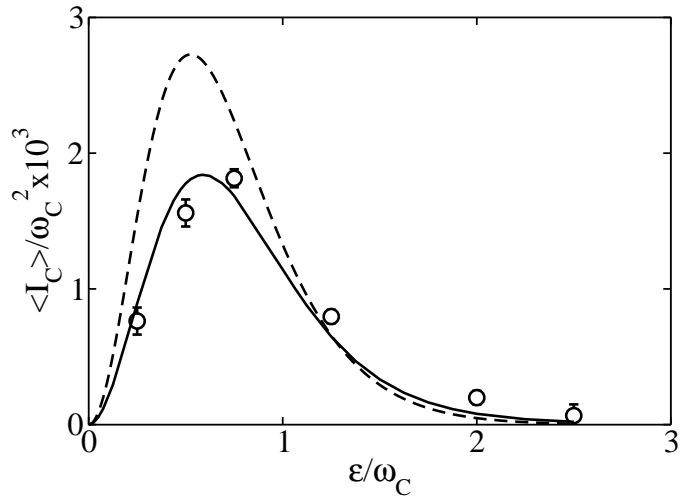}

\vspace{-0.2in}

\end{center}
\caption{\label{HT_TurnOver_eps} Dependence of steady-state heat
current $I_C$ on the TLS energy spacing $\epsilon$.
The results have been obtained with the numerically exact
ML-MCTDH method (circles), the Segal-Nitzan theory (dashed line),
and the NEGF-Redfield approach (full line). The temperatures are
$k_B T_C/\omega_C=0.17$ and $k_B T_H/\omega_H=0.26$ ($\omega_C=\omega_H$).
The TLS-bath coupling strength is $\lambda_C/\omega_C=0.25$ 
($\lambda_C=\lambda_H$). Error bars for some numerically
exact points are smaller than the size of circles, and, therefore, not shown in the figure.}
\vspace{0.2in}
\end{figure}

The influence of the TLS energy spacing $\epsilon$ on the steady-state heat
current in shown in Figure~\ref{HT_TurnOver_eps}.
%--------------------------------------------------------------------
%-----------------------------------------------------------------------
As is seen, the heat current also exhibits a turnover behavior as a function
of the energy spacing. This is due to the resonant character of heat
transport: The heat transport is most efficient if the TLS has the same characteristic
frequency, i.e., energy spacing between the two levels, as that of the baths.
Too large or too small $\epsilon$ (compared to $\omega_c$) both result
in inefficient heat transfer. The detailed discussion of this phenomenon
is given elsewhere.\cite{Velizhanin2008-325}

While the Segal-Nitzan theory can qualitatively predict this turnover behavior,
there is a significant quantitative discrepancy with the numerically exact 
simulation. However, the NEGF-Redfield theory method is in much better agreement
with the numerically exact results in Fig.~\ref{HT_TurnOver_eps}. We attribute
this to the fact that since the coupling strength is relatively small
($\lambda_C/\omega_C=0.25$), a more accurate treatment of TLS-bath coupling 
within the NEGF-Redfield approach leads to a much better agreement with the
numerically exact result.

\section{Concluding Remarks}\label{Concls_sec}

In this paper we have derived an analog of the Meir-Wingreen formula for a
spin-boson nanojunction model for heat transport in a single-molecule junction.
This formula {\em exactly} relates the steady-state heat current through 
a two-level system, connected to two heat reservoirs,
to correlation functions of the operators of the two-level system only.
The formula allows one to calculate the steady-state heat current for the spin-boson
nanojunction model using previously developed methodologies for the
reduced dynamics of the standard spin-boson problem.

As an illustrative application of the formalism developed, we have analyzed the 
properties of the steady-state heat conduction in
the spin-boson nanojunction model
 using the derived expression for the heat current and
Redfield theory to evaluate the correlation functions of the two-level
system.
In addition, a self-consistency criterion was formulated and applied that
enforces the conservation of the heat current within the NEGF-Redfield scheme. 
Employing this approach, we have studied the dependence of the heat current
on the energy spacing of the two level-system and
the coupling strength between the system and the heat baths. The numerical results
obtained in this work demonstrated that the NEGF-Redfield method represents a
significant improvement compared to previous approaches due to the more accurate treatment
of the coupling between the two-level system and the heat baths.

Finally, we would like to comment on the relation of the Meir-Wingreen formula (\ref{SSflux})
developed in this work to previously developed NEGF-based theories of thermal transport.
The latter theories are based on the phonon picture and the anharmonicity, i.e.,
phonon-phonon interactions, is included perturbatively (see Ref.~\onlinecite{Wang2008-381} for an excellent
review and references therein). In contrast, Eq.~(\ref{SSflux}) treats in-bridge
anharmonicity {\em non-perturbatively} since exact eigenstates of an anharmonic
oscillator can serve as bridge levels. For example, the two levels considered in this
may model the two lowest eigenstates of
a double-well potential, which often cannot be treated perturbatively. 
On the other hand, an eigenstate expansion is limited to
a relatively small bridge including only few vibrational modes. Thus, the approach
developed in this paper has to be extended and optimized to apply it to large
molecular bridges. Therefore, the method developed in this work and the
conventional NEGF-based theories of heat transport are independent and complementary.
A major improvement in the theory of heat transport through nanojunctions can thus
be expected if the two methods are combined, allowing the simultaneous treatment of weakly and highly
anharmonic modes of the same nanojunction by anharmonic perturbation theory and
exact diagonalization methods, respectively.

\acknowledgments

K.A.V would like to thank Dima Mozyrsky for helpful discussions on
the diagrammatic technique. M.T. thanks Rainer {H\"artle} for
discussions on the subject of this work. This work has been supported by the National
Science Foundation (NSF) CAREER award CHE-0348956 (H.W., K.A.V), the Deutsche
Froschungsgemeinschaft (DFG) through the DFG-Cluster of Excellence Munich-Centre for Advanced Photonics (M.T.), and used
resources of the National Energy Research Scientific Computing Center (NERSC),
which is supported by the Office of Science of the U.S. Department of Energy
under Contract No. DE-AC02-05CH11231. K.A.V. was also supported by Center for
Nonlinear Studies (CNLS), LANL.

%==========================================================================

%\bibliography{./main}
%\bibliography{main}

\end{document}